\documentclass[%
 aip,
rsi,%
 amsmath,amssymb,
preprint, %
]{revtex4-1}

\usepackage{graphicx}% Include figure files
\usepackage{dcolumn}% Align table columns on decimal point

\usepackage{mathtools,amssymb}   % need for subequations
\usepackage{mathrsfs}

\usepackage{subfig}
\usepackage{siunitx}

\graphicspath{{./figures/}}

\newcommand{\ba}{\begin{eqnarray}}
\newcommand{\ea}{\end{eqnarray}}
\newcommand{\be}{\begin{equation}}
\newcommand{\ee}{\end{equation}}
\newcommand{\eps}{\epsilon}

\usepackage{color}

\begin{document}

\preprint{AIP/123-QED}

\title[Spectral up- and downshifting of Akhmediev breathers under wind forcing]{Spectral up- and downshifting of Akhmediev breathers under wind forcing}% Force line breaks with \\

\author{D. Eeltink}
 \affiliation{GAP-Nonlinear, Universit\'e de Gen\`eve, Carouge, Switzerland}
\author{A. Lemoine}
 \affiliation{Aix-Marseille University, CNRS, Centrale Marseille, IRPHE, Marseille, France}
 \author{H. Branger}
 \affiliation{Aix-Marseille University, CNRS, Centrale Marseille, IRPHE, Marseille, France}
 \author{O. Kimmoun}
 \affiliation{Aix-Marseille University, CNRS, Centrale Marseille, IRPHE, Marseille, France}
 \author{C. Kharif}
 \affiliation{Aix-Marseille University, CNRS, Centrale Marseille, IRPHE, Marseille, France}
 \author{J.~Carter}
 \affiliation{Mathematics Department, Seattle University, Seattle, United States}
  \author{A.~Chabchoub}
 \affiliation{School of Civil Engineering, The University of Sydney, Sydney, Australia}
  \author{ M. Brunetti}
 \affiliation{GAP-Nonlinear, Universit\'e de Gen\`eve, Carouge, Switzerland}
  \author{J. Kasparian} 
\email{jerome.kasparian@unige.ch}
 \affiliation{GAP-Nonlinear, Universit\'e de Gen\`eve, Carouge, Switzerland}

\date{\today}

\begin{abstract}
We experimentally and numerically investigate the effect of wind forcing on the spectral dynamics of Akhmediev breathers, a wave-type known to model the modulation instability. We develop the wind model to the same order in steepness as the higher order modification of the nonlinear Schr\"odinger equation, also referred to as the Dysthe equation. This results in an asymmetric wind term in the higher order, in addition to the leading order wind forcing term. The derived model is in good agreement with laboratory experiments within the range of the facility's length. We show that the leading order forcing term amplifies all frequencies equally and therefore induces only a broadening of the spectrum while the asymmetric higher order term in the model enhances higher frequencies more than lower ones. Thus, the latter term induces a permanent upshift of the spectral mean. On the other hand, in contrast to the direct effect of wind forcing, wind can indirectly lead to frequency downshifts, due to dissipative effects such as wave breaking, or through amplification of the intrinsic spectral asymmetry of the Dysthe equation. Furthermore, the definitions of the up- and downshift in terms of peak- and mean frequencies, that are critical to relate our work to previous results, are highlighted and discussed.
\end{abstract}

\maketitle

% ======= Introduction ====== %
\section{Introduction}
The interaction between wind and ocean waves lies at the heart of ocean dynamics, and since the 1950's \citep{Miles1957,Phillips1957} significant progress has been made in understanding this interplay \citep{Janssen1991, Waseda1999, Miles1996,Leblanc2007,Janssen2009}. More recently, a notable number of studies has been devoted to understanding the occurrence of rogue waves \citep{Pelinovsky2008, Dysthe2008,Akhmediev2009, Onorato2013}. Rogue waves are `monster waves' that have a much higher amplitude than what can statistically be expected from the current sea-state. The role of wind in the formation and the evolution of such extreme waves has been investigated numerically and experimentally \citep{Kharif2008,Chabchoub2013,Toffoli2017}. However, the direct effect of the wind on rogue wave spectra still needs further investigation. An interesting physical phenomenon related to the wind-wave interaction is the frequency downshift observed in ocean waves \citep{Snodgrass1966,Henderson2012}. Apart from altering a fundamental property of a wave field, namely its frequency, downshift can affect the formation of rogue waves \citep{Clamond2006,Islas2011}. Yet, a \textit{unified} explanation for the physical origin of downshift seems to be missing, since proposed mechanisms such as the presence of wind \citep{Hara1991,Touboul2010,Schober2015}, wave breaking \citep{Trulsen1990,Kato1995} and viscosity \citep{Segur2005,Carter2016} are opposite in nature as they respectively force and damp the waves. This study aims to clarify this downshift paradox.

Investigating up- and downshift needs a precise definition. To do so, two approaches can be outlined \citep{Massel1996,Zavadsky2013,Carter2016}. The first is to look at a shift in position of the mode with the highest amplitude in the spectrum, the \textit{spectral peak} $f_\textrm{p}$. In doing so, any small asymmetries in the spectrum are disregarded. The second approach is to look at the position of the \textit{spectral mean} $f_\textrm{m}$, defined as the ratio of the momentum $M$ of the envelope to its norm $N$. In the time domain, these are defined in \citet{Armaroli2017}. In the spectral domain, this can be written as

\begin{align} \label{eqn_defPN}
M &= \sum_{n=-n_\text{lim}}^{n=+n_\text{lim}} f_n |\hat{a}_{f_n}|^2 \\
N  &= \sum_{n=-n_\text{lim}}^{n=+n_\text{lim}}  |\hat{a}_{f_n}|^2 \\
f_m &=  \frac{M}{N} 
\end{align}

\noindent where $\hat{a}$ is the Fourier mode of the envelope, $f$ the frequency, and $n_\text{lim}$ the spectral mode that marks the limit of the main carrier wave mode, such that the interval of $f$ excludes the bound waves. The boundary betweeen the main mode an the first harmonic is considered to be at $f_\text{lim} = f_0 + f_0/2$. From the spectral definition one can see that $M/N$ is equal to the quadratic weighted average and detects any asymmetry in the spectrum. 

In addition, up-/downshifts can be reported as temporary or permanent. In the former case, the mean or the peak shifts to a higher/lower frequency, but eventually shifts back to the original carrier wave frequency. This recovery does not occur in case of permanent up-/downshift.

In this paper, we aim to investigate the effect of the wind on spectra of unidirectional, modulationally unstable gravity wave trains, specifically on Akhmediev breathers. Note that when comparing to the ocean, this excludes directionally spread waves. Our main finding is to inject the wind terms found by \citet{Brunetti2014} into existing MNLS \citep{Dysthe1979} model with viscous dissipation \citep{Carter2016}. This forms a forced-damped MNLS model in which the wind, dissipation and nonlinearity are all of the same order. In section \ref{sec_Model}, we present our model for this purely forcing effect of wind on water waves, that is, wind blowing in the direction of the wave propagation and consequently adding energy to the system. The wind contribution in the model consists of a leading order forcing term that amplifies all wave frequency components equally, as well as a higher order asymmetric term that amplifies higher frequencies more than lower ones. In sections \ref{sec_ExpSimPar} and \ref{sec_ExpSimComp}, we corroborate the model with wave tank experiments. Subsequently, we perform long range simulations in section \ref{sec_LongSim} to overcome the limited fetch in the experiment. We demonstrate that the wind forcing by itself can only cause an upshift in the spectral mean. Finally, in section \ref{sec_Discussion}, we clear up the aforementioned downshift paradox, discussing our model in light of previous literature, and taking into account the two downshift definitions.

% ======= Model ====== %
\section{Weakly nonlinear wind-wave model} \label{sec_Model}
In a simplified Euler approximation for the water-wave problem, the Coriolis term is neglected with respect to the advective term in the momentum equations. In addition, the water density is considered constant $\rho_\textrm{w} = \rho_{\textrm{w,0}}$. Considering unidirectional waves, the transverse coordinate is neglected, and the system of equations that has to be solved is \citep{Dias2008,Kharif2010}:

\begin{align} 
\nabla^2 \phi \equiv \phi_{xx} + \phi_{zz} &= 0 && -h \le z \le \eta(x,t)  \label{eqn_EulerLaplace} \\
\phi_z &= 0 && z=-h \\ 
\eta_t +\phi_x \eta_x - \phi_z &= 2\nu \eta_{xx} && z=\eta(x,t) \label{eqn_Kinematic} \\
\phi_t + \frac{1}{2}\phi_x^2 + \frac{1}{2}\phi_z^2 + g\eta &=  -\frac{P}{\rho_\textrm{w}} -2 \nu  \phi_{zz} && z=\eta(x,t)   \label{eqn_Dynamic}
\end{align}

\noindent where  $g$ is the gravitational acceleration, $h$ is the water depth, $x$ is the propagation direction and $z$ is the upward coordinate. The potential function $\phi = \phi(x,z,t)$ is defined as $u~=~\partial \phi/\partial x~= ~\phi_x$, $w =  \partial \phi/\partial z= \phi_z$, where $u$ and $w$ are the velocity components. The function $\eta(x,t)$ is the surface elevation with respect to the average level $z=0$. The kinematic viscosity of the fluid is denoted as $\nu$ [m$\textsuperscript{2}$/s]. The wind is characterized by $P=P(x,t)$, the excess pressure at the water surface in the presence of wind, that in the framework of the Miles mechanism \citep{Miles1957} is given by
\be
\frac{P}{\rho_\textrm{w}} = \frac{\Gamma}{f_0} \frac{c_p^2}{2\pi} \eta_x =  \frac{\omega_0}{k_0^2} \Gamma \eta_x 
\ee

\noindent where $c_p$ is the phase velocity of the carrier wave, $k_0$ its wave number, $f_0$ its frequency, and  $\omega_0 = 2 \pi f_0$ its angular frequency. $\Gamma$ [1/s] is the growth rate of the energy $E$ of the waves due to the wind blowing in the direction of the wave propagation, that is $\frac{\partial E}{\partial t} = \Gamma E$. In turn, an expression for $\Gamma$ as a function of wind speed $U$ and wave frequency $f_0$ can be modeled in various ways, as will be discussed in section \ref{sec_ExpSimPar}. In the approximation that the envelope varies slowly in comparison to the surface elevation, the Method of Multiple Scales (MMS) can be used to expand these boundary conditions and obtain a weakly nonlinear propagation equation for the envelope $a(x,t)$ at each order of interest. In the MMS, the small order parameter is the steepness of the wave $\eps = ak_0$. In addition, deep water waves, $k_0 h \rightarrow \infty$, are considered. The surface elevation to the first order in steepness is given by
\be 
\label{eqn_env}
\eta (x, t) = \operatorname{Re}\{ a(x, t) \exp[i (k_0 x - \omega_0 t)] \}
\ee

Note that the sign choice of  the argument in the exponential is important for the spectral representation of envelope with respect to the carrier wave. For instance in \citet{Carter2016} the opposite sign choice has been made. To avoid confusion, the spectrum of the carrier wave is plotted throughout this paper. In the absence of both wind and viscosity ($\nu =  \Gamma = 0$), the MMS yields (i) the nonlinear Schr\"{o}dinger (NLS) equation when developing the boundary conditions to $O(\eps^3)$ \citep{Hasimoto1972, Mei2005, Ablowitz2011}, and (ii) the modified NLS (MNLS) or Dysthe equation when developing the boundary conditions to $O(\eps^4)$ \citep{Dysthe1979}. Assuming the wind $\Gamma/f_0$ and the viscosity $\nu k_0^2/f_0$ contribution are both of order $O (\epsilon^2)$, and including these in the MMS development up to $O(\eps^4)$ using a similar method as \citet{Carter2016}, yields the following damped-forced MNLS:
\be\label{eqn_MMSOutput}
\begin{split}
 \frac{\partial a}{\partial t} + \frac{\omega_0}{2k_0}  \frac{\partial a}{\partial x} =& \eps\bigg[ -i\frac{\omega_0}{8 k_0^2}  \frac{\partial^2 a}{\partial x^2} - \frac{1}{2}i k_0^2 \omega_0 a|a|^2 -2 k_0^2\nu a  + \frac{1}{2}\Gamma a\bigg] \\
& + \eps^2\bigg[ 4i k_0 \nu\frac{\partial a}{\partial x}  -  \frac{3i}{4 k_0}\Gamma \frac{\partial a}{\partial x}\\
 &- \frac{3}{2} k_0 \omega_0 |a|^2\frac{\partial a}{\partial x} - \frac{1}{4} k_0 \omega_0 a^2 \frac{\partial a^*}{\partial x} + \frac{\omega_0}{16k_0^3}\frac{\partial^3 a}{\partial x^3}-i k_0 a\frac{\partial \bar{\phi}}{\partial x}\bigg]
 \end{split}
\ee

\noindent where $\bar{\phi}$ is the potential mean flow. To obtain a propagation in space, in accordance with the motion in a 1D wave tank, the following coordinate transformation is applied:
\[
    \left\{
                \begin{array}{ll}
                  \tilde{t} = t - \frac{2k_0}{\omega_0} x\\
                  \tilde{x} = \eps  x\\
                \end{array}
              \right.
\]

The potential mean flow term in (\ref{eqn_MMSOutput}) is replaced by a Hilbert transform term, see \citet{Janssen1983}. The Hilbert transform $\mathscr{H}$ is defined as $\mathscr{F}[\mathscr{H}[u]]=-i~\operatorname{sign}(\omega)\mathscr{F}[u]$, where $\mathscr{F}$ is the Fourier transform. The damped-forced MNLS becomes:

\be\label{eqn_FullModel}
\begin{split}
\underbrace{ \frac{\partial a}{\partial \tilde{x}} + i\frac{k_0}{\omega_0^2}  \frac{\partial^2 a}{\partial \tilde{t}^2} + i{k_0^3} a|a|^2}_\textrm{NLS} =  & \underbrace{+  \eps\frac{k_0^3}{\omega_0}\bigg(6 |a|^2\frac{\partial a}{\partial \tilde{t}} + 2 a\frac{\partial |a|^2}{\partial \tilde{t}} + 2ia \mathscr{H}\left[\frac{\partial |a|^2}{\partial \tilde{t}}\right]\bigg)}_\textrm{MNLS correction} \\ 
&  \underbrace{-4 \frac{k_0^3}{\omega_0}\nu a-\eps 20 i \frac{k_0^3}{\omega_0^2}\nu\frac{\partial a}{\partial \tilde{t}}}_\textrm{Viscosity} \\ 
& \underbrace{+ \frac{k_0}{\omega_0}\Gamma a  + \eps4i \frac{k_0}{\omega_0^2}\Gamma\frac{\partial a}{\partial \tilde{t}}}_\textrm{Wind} \\ 
 \end{split}
\ee

This full model consists of the corrected NLS, taking into account higher-order dispersion and mean flow (MNLS correction) as well as viscosity and wind effects, where the higher order terms are indicated by $\eps$. In the following the tilde's are omitted. The dynamics of the MNLS equation are well known and have been studied numerically by \citet{Lo1985,Adcock2016} and applied to many experiments, see for instance \citet{Trulsen2001,Tulin1999,Slunyaev2013,Chabchoub2013a}. 

The viscosity consists of a linear damping term $-4 \frac{k_0^3}{\omega_0}\nu a$ in the leading order, as proposed in \citet{Dias2008}. In addition, following \citet{Carter2016}, the higher order viscosity term is given by $-\eps 20 i \frac{k_0^3}{\omega_0^2}\nu\frac{\partial a}{\partial t}$. 

The wind action appears as a leading order linear forcing term $\frac{k_0}{\omega_0}\Gamma a$ which in past work has been included based on either the intuition of a simple forcing \citep{Trulsen1992}, or through more rigorous justification \citep{Kharif2010}. There has been no experimental validation of a wind forcing term in an evolution equation for deep water waves. Our addition of a higher order wind term $\eps4i \frac{k_0}{\omega_0^2}\Gamma\frac{\partial a}{\partial t}$ completes the wind-wave model (\ref{eqn_FullModel}). This higher order wind term restores consistency in the sense that since the nonlinearity is developed up to $O (\eps^4)$ in the MNLS framework, the dissipation and forcing also have to be developed up to this order to have a coherent model. 

The model reported here can be related to the wind-wave model presented in \citet{Brunetti2014a,Brunetti2014} by noting that the higher order wind term can be obtained in two ways. To obtain (\ref{eqn_FullModel}), the wind forcing $\Gamma/f_0$ was assumed $O(\eps^2)$ in (\ref{eqn_Dynamic}), and the kinematic and dynamic boundary conditions were developed up to $O (\eps^4)$ in the MMS. Alternatively, following \citet{Brunetti2014a}, one can assume that $\Gamma/f_0 = O(\eps)$ in (\ref{eqn_Dynamic}), thus, strong in comparison to the steepness $\eps$. Next, the kinematic and dynamic boundary conditions only have to be developed up to $O (\eps^3)$ in the MMS to obtain the same wind terms as in (\ref{eqn_FullModel}).

Surprisingly, the outcome of the MMS is that the viscosity and wind contributions in both the leading and higher order have the same form in (\ref{eqn_FullModel}), despite of their non-similar appearance in the  kinematic and dynamic boundary conditions ( \ref{eqn_Kinematic} and \ref{eqn_Dynamic}), respectively. As these terms have opposite signs, the viscosity can cancel out the wind forcing, and vice versa. However, note that their balance is different for different orders. This balance occurs as 
\be \label{eqn_delta_0}
\delta_0  \equiv \Gamma-4 k_0^2 \nu = 0
\ee

\noindent for the leading order, as studied by  \citet{Kharif2010}. In our model (\ref{eqn_FullModel}), the higher order terms are balanced if 
\be \label{eqn_delta_1}
\delta_1 \equiv \Gamma - 5 k_0^2 \nu = 0
\ee

The NLS is a spectrally symmetric evolution equation, that is, as the envelope propagates, the frequencies below $f_0$ evolve in the same way as their counterparts above $f_0$ \citep{Segur2005}. An odd derivative in time is needed to have an asymmetric evolution  of the spectrum \citep{Lo1985}. In the full model (\ref{eqn_FullModel}), the odd terms are the MNLS correction, the higher order wind and the higher order viscosity term. Indeed, the leading order term $\delta_0$ has the effect of either damping or amplifying all frequencies in the spectrum, depending on its sign. For the higher order, if viscosity is dominant, $\delta_1 < 0$, the higher frequencies ($f>f_0$) are damped more than lower ones ($f<f_0$) causing a permanent downshift of the spectral mean, as evidenced by \citet{Carter2016}. Our derivation of the higher order wind term shows that if the wind action is dominant, $\delta_1 > 0$, the higher order effect is opposite and the spectral mean is upshifted.

% ======= Experiment and Simulations Details ====== %
\section{Experimental and computational details} \label{sec_ExpSimPar}

% --  Initial Condition -- %
\subsection{Initial condition}

To validate our model, we are interested in cases in which the system experiences a spectral broadening. To model the modulation instability, the Akhmediev Breather (AB) \citep{Akhmediev1985} has been used to generate our initial conditions for the experiment. Starting the dynamics from an exact AB expression is useful for experimental investigations, since it allows to trigger the modulation instability dynamics in relation to the length of the facility. The AB gives an approximate prediction of the growth and subsequent decay cycle that is to occur, rather than just a prediction of a linear growth rate as calculated from the Benjamin-Feir instability analysis \citep{Benjamin1967}. This theoretical expectation allows us to identify deviations from this growth-decay pattern. The AB is a solution to the $NLS$ part of (\ref{eqn_FullModel}) and reads \citep{,Kibler2010} 

\be \label{eqn_AB}
a_\text{AB} (x^*,t) = a_0\frac{\sqrt{2A} \cos(\Omega \frac{t}{T_0}) + (1-4A) \cosh(R \frac{x^*}{L0}) + iR \sinh(R \frac{x^*}{L_0})}{\sqrt{2A} \cos(\Omega \frac{t}{T_0}) - \cosh(R \frac{x^*}{L_0})} \exp(i \frac{x^*}{L_0}),
\ee

\noindent where $x^*$ is the distance to the focal point, $L_0 = (k_0^3 a_0^2)^{-1}$ and $T_0 = \sqrt{2}(k_0\omega_0 a_0)^{-1}$ are the rescaling coordinates, $\Omega = 2 \sqrt{1 - 2A}$ is the dimensionless modulation frequency, $R~=~\sqrt{8A(1~-~2A)}$ the growth rate, and $0 < A < 0.5$ is the Akhmediev parameter. The case $A = 0.25$  corresponds to a maximal growth rate in the Benjamin-Feir theory. Note that since the AB is a solution of the NLS only, the maximally unstable mode in the NLS framework does not exactly coincide with that of our full system. When $x \rightarrow \pm\infty$ the AB tends to a regular wave train, while at the focal point $x = 0$, the breather reaches maximal temporal compression and consequently maximal amplitude. In the spectral domain, the focusing of the breather corresponds to a broadening of the spectrum, see \citet{Wetzel2011}. Thus, the focal point is marked by maximal spectral width, and by minimal amplitude of the carrier wave Fourier component.

Choosing an initial condition at a given distance from the focal point allows us to control the number of developed sidebands. The signal given to the wave maker is the surface elevation $\eta (x^*,t)$  (\ref{eqn_env}), based on the dimensional AB envelope $a_\text{AB}(x^*,t)$ (\ref{eqn_AB}).

% -- Experimental Setup -- %
\subsection{Experimental Setup}

% Figure Lumminy %
\begin{figure}  
        \includegraphics[width=0.8\textwidth]{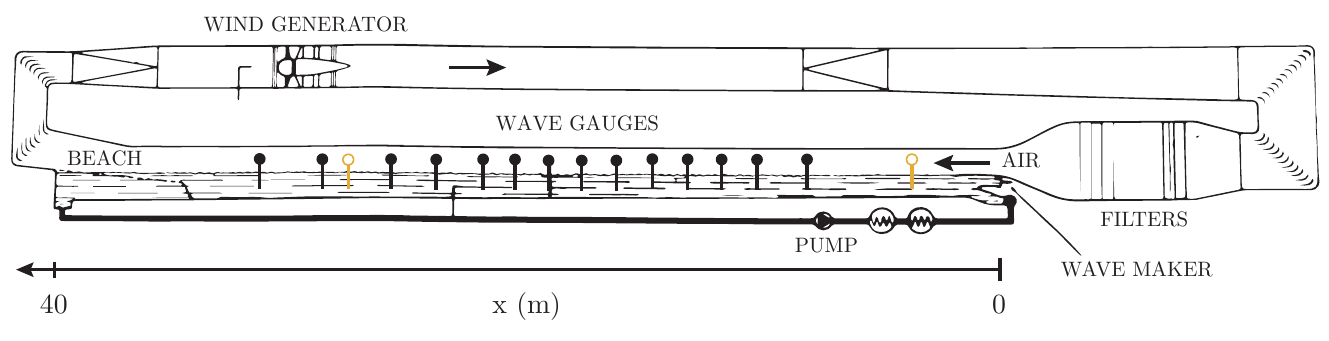}
     \caption{IRPHE wind-wave tank facility: 40 m long, 3 m wide, water depth 0.80 m, air channel height 1.5 m. Wave gauges have been placed approximately evenly along the tank. Capacitive wave gauges are indicated with open circles, others are resistive gauges}
     \label{fig_luminy}
\end{figure}
 
Experiments have been performed in the closed wind-wave facility at IRPHE (Luminy)-Aix Marseille University. A schematic depiction of the facility is shown in figure \ref{fig_luminy}. As detailed in \citet{Coantic1981}, the tank is 3 m wide, has a water depth of 80 cm, and a length of 40 m. At the end of the wave tank, an 8~m sloping beach was installed to prevent wave reflection. At the beginning of the tank, a 1.5 m long plastic sheet floating on the water surface allowed the incoming wind to be tangential to the water surface, and damped possible high-frequency mechanical wave modes. Mechanical waves have been generated by an underwater piston-like wavemaker controlled by a computer. The system was able to produce arbitrary surface gravity waves in the frequency range of 0.5-1.8 Hz. The air-channel above the tank is 1.5 m high. The wind was generated by a closed-loop air flow system, up to a maximum velocity of $U$ = 15 m/s in the direction of the wave propagation. A total of 16 wave gauges have been placed at fixed positions along the tank to measure the surface elevation. The gauges were positioned approximately evenly between $x$ = 3 m and $x$ = 32 m. The first and 14\textsuperscript{th} gauge were capacitive-, the others were resistive-type wave probes. All gauges had a sampling rate of 400 Hz. The wind speed was measured by a pitot tube at different positions in the tank to verify a constant and homogeneous air flow. 

The parameter space of the experiment is limited in several directions. Firstly, the steepness of the background plane wave of the AB (when $x^*  \rightarrow -\infty$) is constrained to a range of $0.08 \leq \eps \leq 0.10$. The lower the steepness, the larger the propagation distance required for the modulation instability to develop in the small fetch, restricted by the size of the facility (figure \ref{fig_luminy}). Conversely, if the carrier waves are initially too steep, wave breaking is inevitable as a consequence of focusing. Secondly, all measurements have been performed with $f_0$ = 1.67 Hz, thus $\omega_0$ = 10.5 rad Hz, and $k_0$ =~ 11.2 rad/m. This yields a Bond number of $\sim$1000, confirming these waves are in the gravity wave regime \citep{Dias1999}. Higher frequencies waves were not possible to generate with the installed wavemaker. Thirdly, wind speeds were limited, as for $U \gtrsim$ 4 m/s breaking would occur for waves generated with initial steepness in the described range. All results presented are free of energetic wave breaking, unless specifically mentioned otherwise.

Besides the breather-type waves generated by the wavemaker, the wind naturally generates additional waves. The frequency range of these wind-waves shifts down as a function of wind speed and fetch. At the last wave gauge, for $U$~=~4~m/s, the wind-waves have been measured to be in the range of approximately 2.2 - 3.2 Hz. Therefore, these waves were considered not to overlap with the breather-type waves. The wind-generated waves showed micro-breaking at wind speeds of $U \sim$~4~m/s towards the end of the tank.

% -- Simulation Parameters -- %
\subsection{Simulation parameters}

% FIGURE: Initial Condition Waveform%
\begin{figure}  

    \centering
    \begin{subfloat}[\label{fig_ICwave}]
        {\includegraphics[width=0.8\textwidth]{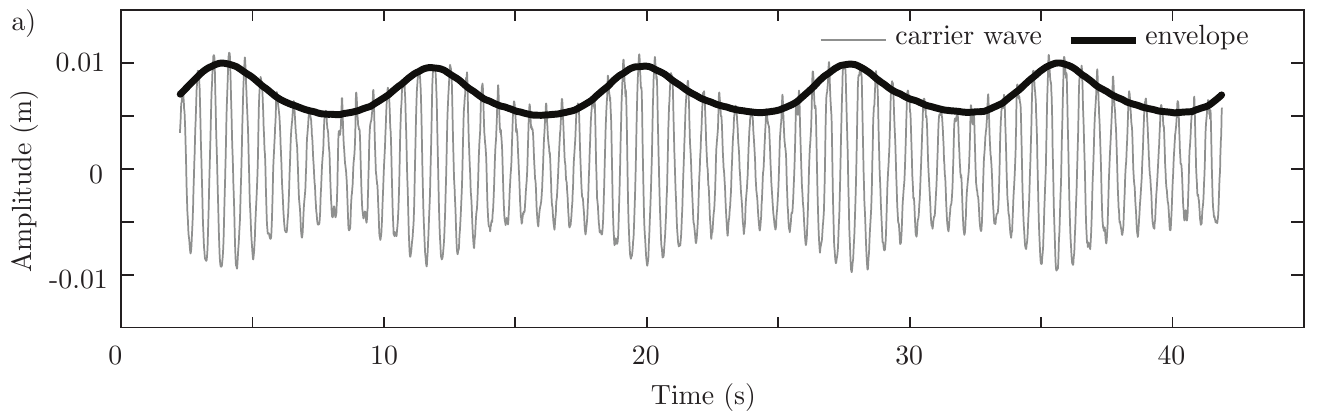}}
    \end{subfloat}
    \begin{subfloat}[\label{fig_IC_spec_exp}]
        {\includegraphics[width=0.4\textwidth]{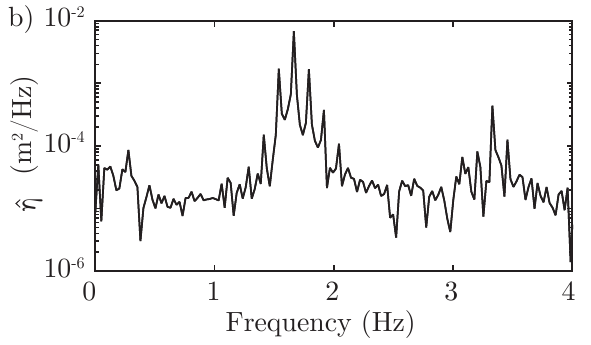}}
    \end{subfloat}
   \begin{subfloat}[\label{fig_IC_spec_sim}]
        {\includegraphics[width=0.4\textwidth]{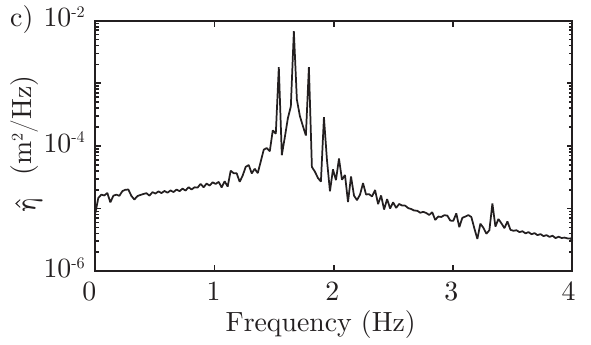}}
    \end{subfloat}
     \caption{a) Periodic initial condition used for the simulations, as measured by the first wave gauge at $x$ = 3.7 m at an acquisition rate of 400 Hz. b) Log-linear plot of the spectrum of the surface elevation as measured by the first wave gauge. c) Log-linear plot of the spectrum of the carrier wave to first order as calculated from the envelope. Carrier wave parameters $\eps$ = 0.08, $T$ = 0.6 s ; AB parameters $x^*=-30$ m and $A$ = 0.25. The envelope was calculated using the Hilbert transform.}
     \label{fig_IC}
\end{figure}

The envelope $a$, which is periodic in time, was integrated forward in space according to (\ref{eqn_FullModel}) by means of a split-step Fourier scheme \citep{Hardin1972}. As described in \citet{Agrawal2001}, the linear and nonlinear part of (\ref{eqn_FullModel}) can be solved in separate steps. The linear part is an ordinary differential equation (ODE) which was solved in Fourier space, and to integrate the nonlinear part, the fourth order Runge-Kutta method has been used. 

This initial condition for the simulations is the envelope based on the surface elevation measured by the first wave gauge. To ensure periodic boundary conditions, we selected a time interval equal to a multiple of the envelope period (see figure \ref{fig_IC}). As such, the spectral resolution of the simulation is the same as that in the experiment. At an acquisition rate of 400 Hz, depending on the exact conditions, the time series were $\sim$ 40 s long, corresponding to $\sim$ 60 periods of the carrier wave and spectral resolution of $\sim$ 0.025 Hz. The complex envelope was extracted from the real valued envelope $\eta$ as follows

\be
a (x_0,t) = \left[\eta(x_0,t) + i \tilde{\eta}(x_0,t) \right]e^{-i(k_0 x-w_0 t)} 
\ee

\noindent where $\tilde{\eta}$ is the Hilbert transform of the surface elevation, as described in \citet{Osborne2010} and $x_0$ = 3.1 m is the position of the first wave gauge. Starting the simulations from the experimental signal accounts for the possible imperfections of the wavemaker and reduces the number of free parameters in the model. The only free parameters to be determined are the wind growth rate $\Gamma$ and the viscosity parameter $\nu$, both of which are fitted from the experiment.

Following the method of \citet{Segur2005} and \citet{Carter2016}, the viscosity parameter $\nu$ in (\ref{eqn_FullModel}) is in fact an effective term that includes not only viscosity, but all sources of dissipation in the experiment such as the side-wall effects and surface contamination. It has been determined experimentally by fitting the exponential decay of a the wave train propagating down the tank without wind forcing: $E = E_0 e^{- \frac{1}{2}\zeta x}$, where $\zeta = (4 k_0^3/\omega_0) \nu$. All waves used in this work are of the same frequency and steepness, for which we found a measured value of the viscosity of $\nu~=~\SI[separate-uncertainty = true]{1.18(35)e-5}{m^2/s}$ . Consequently, we used $\nu~=~\SI{1e-5}{m^2/s}$ in all simulations.

% ---- Table Gamma ---- %
\begin{table*}
\caption{\label{tab_Gamma}Predicted values and values used in the simulations for the wind growth rate $\Gamma$}
\begin{ruledtabular}
  \begin{tabular}{l@{\hskip 0.2in}c@{\hskip 0.2in}c@{\hskip 0.2in}c@{\hskip 0.2in}c}
     Wind speed & $\Gamma_\text{Miles}$   &   $\Gamma_\text{Simulation}$  &  leading order  &  higher order   \\[3pt] \hline
       4 m/s  & $\SI{8.3e-3}{s^{-1}}$ & $\SI{8.5e-3}{s^{-1}} $ & $\delta_0 > 0 $ & $\delta_1 > 0 $ \\
       2 m/s & $\SI{1.6e-3}{s^{-1}}$ & $\SI{2.0e-3}{s^{-1}} $  & $\delta_0 < 0 $ & $\delta_1 < 0 $  \\
  \end{tabular}
\end{ruledtabular}
\end{table*}

To determine $\Gamma$ we rely on predictions by the Miles mechanism for wind growth \citep{Banner2002a}
\be
\Gamma = \omega_0 \alpha \frac{\rho_{air}}{\rho_w} \frac{u_*^2}{c_p^2} 
\ee

\noindent where $u_*$ is the friction velocity, $c_p$ is the phase velocity,  $\rho_{air}$ the density of air, $\rho_w$ the density of water, and $\alpha$ an empirical parameter of 32.5. Assuming a logarithmic profile of $U$ as a function of $z$, the measured wind speed $U$ is related to the friction velocity $u_*$ by \citep{Miles1957}
\be
U (z) = \frac{u_*}{K} \log \left(\frac{z}{z_0}\right) , 
\label{eqn_U}
\ee
where $z$ is the height where $U$ is measured, $K = 0.41$ the Von Karman constant and $z_0$ is the effective roughness length
\be
z_0 = \kappa u_*^2/g  ,
\ee
with $\kappa = 0.0144$ the Charnock constant.

Since estimations for $\Gamma$ can have deviations up to a factor two with field and wave tank measurements (figure 1 in \citet{Banner2002a}), we adjusted the latter value to match the spectral widening and growth in the experiment for each wind speed, see table \ref{tab_Gamma}. Values obtained for $u_*$ through (\ref{eqn_U}) are accurate within a range of 10 \% compared to those measured in an elaborate study by \citet{Caulliez2008a} in the same facility.

% ==== Experimental results and model validation ===== %
\section{Experimental results and model validation} \label{sec_ExpSimComp}

Our numerical wind-wave model was compared to laboratory experiments on AB-type waves. All results presented here are based on carrier wave parameters $T~=~\frac{2 \pi}{\omega_0}~=~0.6$~s ($f_0~=~1.67$ Hz), $\eps$~=~0.08 and AB parameter $A$ = 0.25, yielding the modulation frequency $\Delta f~=~0.13$ Hz. For the simulations, we set $\nu~=~\SI{1e-5}{m^2/s}$ and $\Gamma$ as presented in table \ref{tab_Gamma}.

% Figure Spectra %
\begin{figure}
    \centering
    \begin{subfloat}[\label{fig_CompExp_Xf30_W0_spect_exp}]
        {\includegraphics[width=0.3\textwidth]{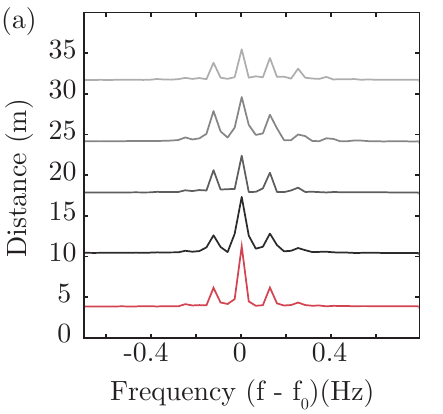}}
    \end{subfloat}
    \begin{subfloat}[\label{fig_CompExp_Xf30_W0_spect_sim}]
        {\includegraphics[width=0.3\textwidth]{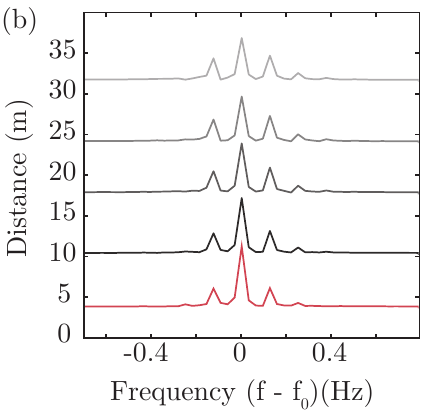}}
    \end{subfloat}
    \begin{subfloat}[\label{fig_CompExp_Xf30_W4_spect_exp}]
        {\includegraphics[width=0.3\textwidth]{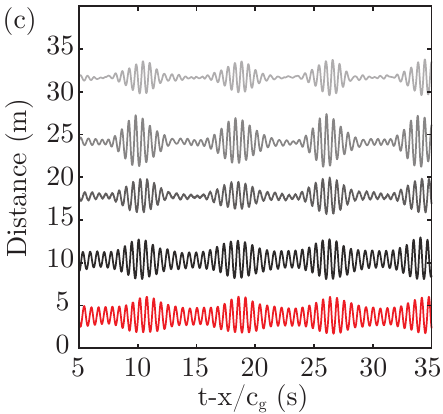}}
    \end{subfloat}
    \begin{subfloat}[\label{fig_CompExp_Xf30_W4_spect_sim}]
        {\includegraphics[width=0.3\textwidth]{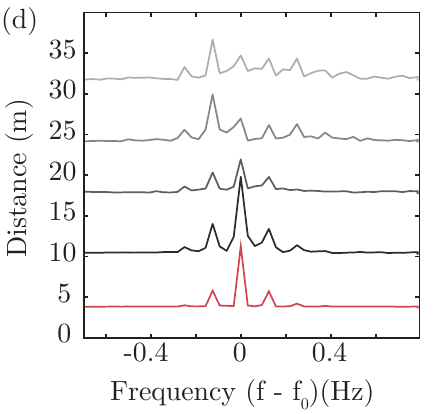}}
    \end{subfloat}
    \begin{subfloat}[\label{fig_CompExp_Xf30_W2_eta}]
        {\includegraphics[width=0.3\textwidth]{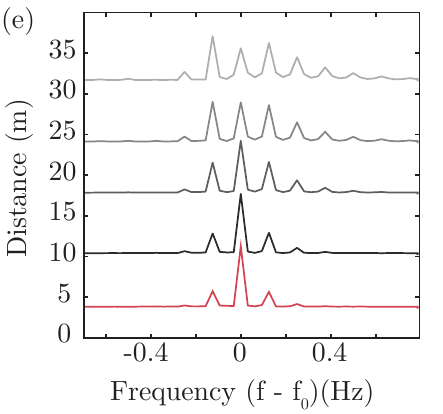}}
    \end{subfloat}
    \begin{subfloat}[\label{fig_CompExp_Xf30_W4_eta}]
        {\includegraphics[width=0.3\textwidth]{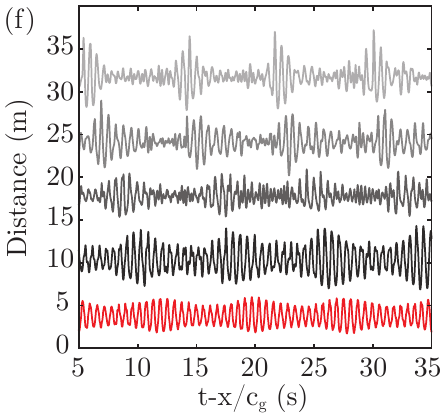}}
    \end{subfloat}
     \caption{Spectral amplitude $|\hat{\eta}|$ (normalized) offset by distance, for carrier wave parameters $\eps$ = 0.08, $T_0$ = 0.6 s, AB parameters $x^*=-30$ m and $A$ = 0.25, and simulation parameter $\nu = \SI{1e-5 }{m^2/s}$. (a,d) Experimental spectral evolution, (b,e) corresponding simulations. (c,f) Experimental surface elevation evolution. }(a,b,c) $U$ = 0 m/s, (d,e,f) $U$ = 4 m/s. Initial condition (simulation) and measurement (experiment) at $x$ = 3.1 m are indicated by the red line. \label{fig_CompExp_Xf30_Spect}
\end{figure}

Based on the analytical solution to the NLS, it is expected for the AB to reach its maximal focusing after propagating 30 m from the wave maker ($x^*=-30$~m). Figure \ref{fig_CompExp_Xf30_Spect} compares the evolution of the spectra retrieved from the experiment to the numerical simulations, without (a,b) and with the presence of wind $U$ = 4 m/s (c,d). Note that the initial spectrum is still symmetric, both with and without wind, i.e. the lower sideband $\hat{\eta}_{-1}$ and the upper sideband $\hat{\eta}_{+1}$ have a similar amplitude. In the experiment without wind, see figure \ref{fig_CompExp_Xf30_W0_spect_exp}, due to the MNLS correction \citep{Trulsen2001,Slunyaev2013}, and taking into account the effect of viscosity, the focal point is expected to be around $x=36$~m. Note that the higher modes $\hat{\eta}_{+2}$ and $\hat{\eta}_{+3}$ slightly grow towards the end of the tank. Indeed, our simulations in figure \ref{fig_CompExp_Xf30_W0_spect_sim} reproduce these features. In the presence of wind, $U$ =~4~m/s, both the experimental measurements in figure \ref{fig_CompExp_Xf30_W4_spect_exp} and the simulations in figure \ref{fig_CompExp_Xf30_W4_spect_sim} show an upstream shifting of the focal point, that is, $\hat{\eta}_0$ reaches a minimum around $x=28$~m. In addition, under the wind action, the spectrum broadens, albeit in an asymmetric manner as a broad range of higher frequencies $\hat{\eta} > \hat{\eta_0}$ grow, while only a narrow range of the lower frequencies $\hat{\eta} < \hat{\eta_0}$ do. 

\begin{figure}
    \centering
    \begin{subfloat}[\label{fig_CompExp_Xf30_w0_sideband_exp2}]
        {\includegraphics[width=0.3\textwidth]{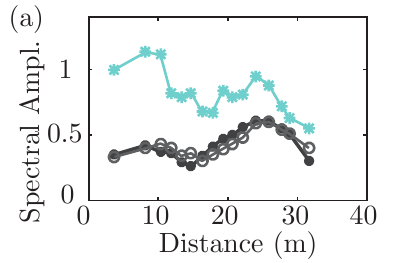}}
    \end{subfloat}
    \begin{subfloat}[\label{fig_CompExp_Xf30_w0_sideband_sim2}]
        {\includegraphics[width=0.3\textwidth]{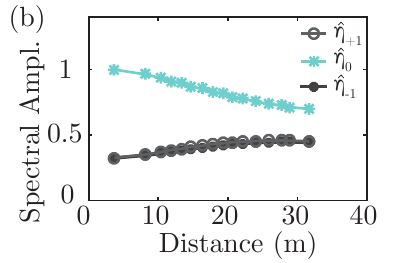}}
    \end{subfloat}
    \begin{subfloat}[\label{fig_CompExp_Xf30_w0_E}]
        {\includegraphics[width=0.3\textwidth]{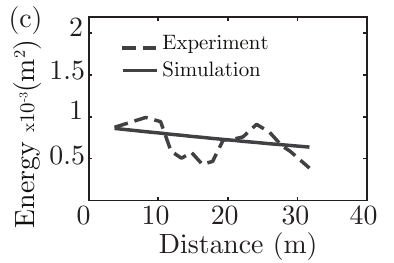}}
    \end{subfloat}
     \begin{subfloat}[\label{fig_CompExp_Xf30_w2_sideband_exp2}]
        {\includegraphics[width=0.3\textwidth]{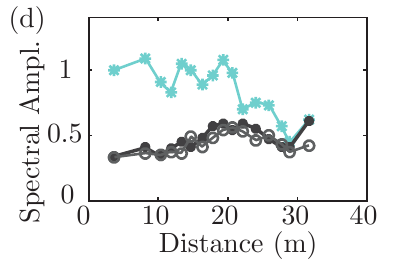}}      
    \end{subfloat}
    \begin{subfloat}[\label{fig_CompExp_Xf30_w2_sideband_sim2}]
        {\includegraphics[width=0.3\textwidth]{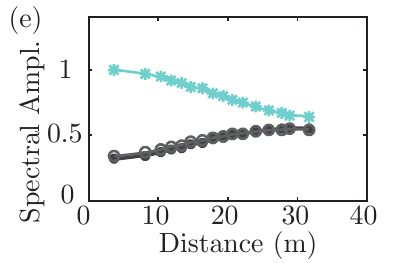}}      
    \end{subfloat}
        \begin{subfloat}[\label{fig_CompExp_Xf30_w2_E}]
        {\includegraphics[width=0.3\textwidth]{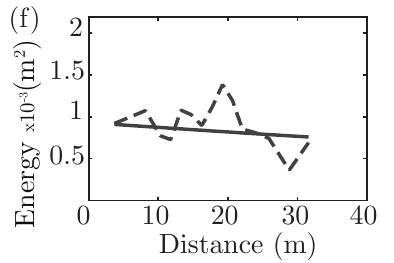}}
    \end{subfloat}
     \begin{subfloat}[\label{fig_CompExp_Xf30_w4_sideband_exp2}]
        {\includegraphics[width=0.3\textwidth]{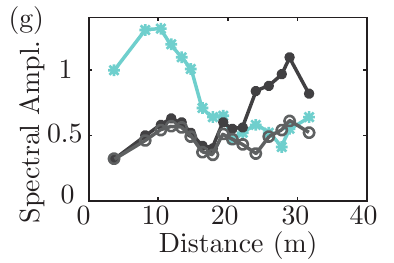}}    
    \end{subfloat}
    \begin{subfloat}[\label{fig_CompExp_Xf30_w4_sideband_sim2}]
        {\includegraphics[width=0.3\textwidth]{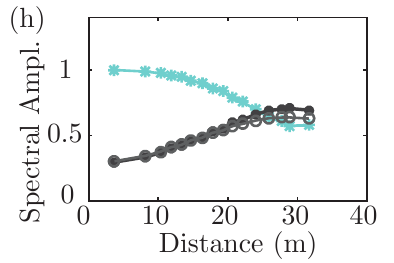}}     
    \end{subfloat}
            \begin{subfloat}[\label{fig_CompExp_Xf30_w4_E}]
        {\includegraphics[width=0.3\textwidth]{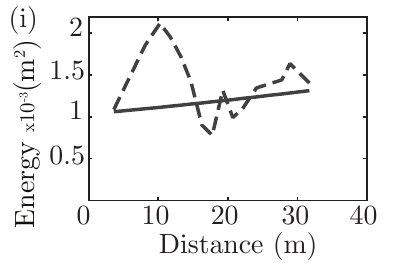}}
    \end{subfloat}
     \caption{Evolution of the three central Fourier components $\hat{\eta}_0$ ($\star$),  $\hat{\eta}_{+1}$ ($\circ$) and  $\hat{\eta}_{-1}$ ($\bullet$), for the same parameters as in figure \ref{fig_CompExp_Xf30_Spect}. (a,d,g) Experiments, (b,e,h) simulations. Values are normalized to the amplitude of $\hat{\eta}_0$ at the first wave gauge. (c,f,i) Energy evolution. (a,b,c) $U$~=~0~m/s, (d,e,f) $U$~=~2~m/s, (g,h,i) $U$ = 4 m/s. }\label{fig_CompExp_Xf30_sideband}
\end{figure}

In order to characterize the appearance of the spectral asymmetry, and see the effect of the wind on the shifting of the focal point, figure \ref{fig_CompExp_Xf30_sideband} displays the evolution of the carrier wave Fourier component $\hat{\eta}_0$  as well as the first upper and lower side bands as a function of propagation distance. With increasing wind speed the decay rate for $\hat{\eta_0}$ increases, equally, the growth rate of $\hat{\eta}_{-1}$ and $\hat{\eta}_{+1}$ increases. Consequently, the crossing point, at which the amplitude of the sidebands overtake $\hat{\eta}_0$, moves upstream with increasing wind speed. For $U$ = 4 m/s, since $\hat{\eta}_{-1}$ has a higher growth rate than $\hat{\eta}_{+1}$, a downshift of the spectral peak originates, as measured at the last wave gauge. The numerical simulations of our developed model reproduce this shifting behavior. Similarly, the simulations matches the trend for the evolution of the total energy, in spite of the experimental fluctuations due to the inherent variability related to wind. Note that due to the width and finite length of the tank, a certain amount of this fluctuation in total energy measured by each wave gauge can also vary due to for instance wave reflections on the sloping beach at the end of the tank and the lateral alignment of the wave gauges, causing fluctuations superimposed on the trend set by the wind and viscosity input. The spectrum however, is not affected by these factors.

\begin{figure}
    \centering
    \begin{subfloat}[\label{fig_CompExp_Xf30_w0_peakmean_exp}]
        {\includegraphics[width=0.4\textwidth]{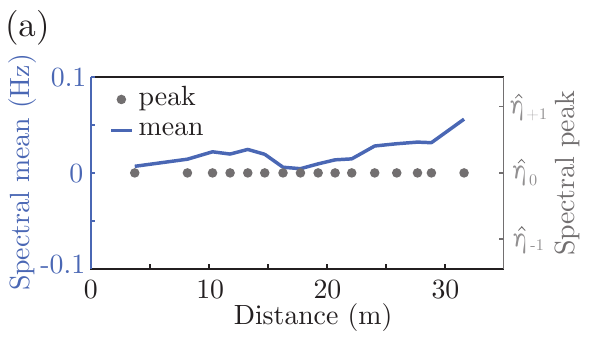}}      
    \end{subfloat}
    \begin{subfloat}[\label{fig_CompExp_Xf30_w0_peakmean_sim}]
        {\includegraphics[width=0.4\textwidth]{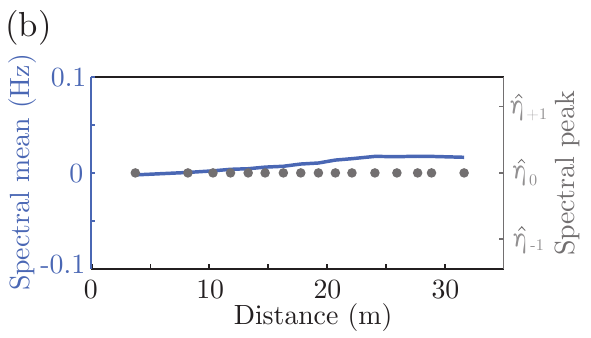}}     
    \end{subfloat}
    \begin{subfloat}[\label{fig_CompExp_Xf30_meanmax_exp}]
        {\includegraphics[width=0.4\textwidth]{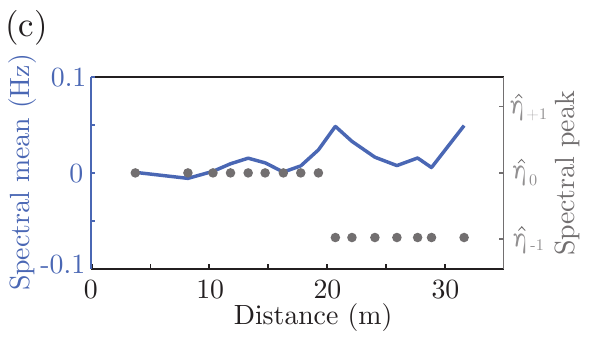}}       
    \end{subfloat}
    \begin{subfloat}[\label{fig_CompExp_Xf30_meanmax_sim}]
        {\includegraphics[width=0.4\textwidth]{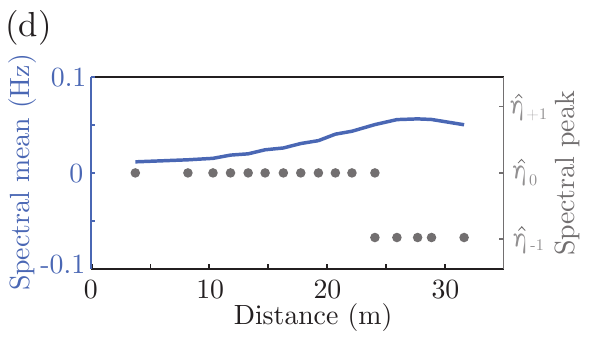}}       
    \end{subfloat}
     \caption{Evolution of the spectral mean $f_\textrm{m}$ (blue solid line, left vertical axis) and spectral peak $f_\textrm{p}$ (gray dots, right vertical axis), for same parameters as in figure \ref{fig_CompExp_Xf30_Spect}. (a,c) Experiments, (b,d) simulations. (a,b) $U$ = 0 m/s, (c,d) $U$ = 4 m/s. Two dots at the same distance indicates the spectral heights are within 1 percent range.}\label{fig_CompExp_Xf30_MeanPeak}
\end{figure}

Figure \ref{fig_CompExp_Xf30_MeanPeak} shows the evolution of spectral peak $f_\textrm{p}$ (dots) and the spectral mean $f_\textrm{m}$ (solid line) as a function of distance. The spectral peak remains equal to $f_0$ for $U$ = 0 m/s, while for $U$ = 4 m/s a downshift of the peak occurs. Due to the length limitations of the tank and the short considered fetch, we are not able to experimentally assess and state whether this shift is permanent or temporary. In contrast to the spectral peak, the spectral mean in figure \ref{fig_CompExp_Xf30_MeanPeak} demonstrates a clear upshift. This dissimilar behavior illustrates the need for a careful definition of up- and downshift in order to allow for an accurate description of the physics and dynamics at play.

\begin{figure}
    \centering
    \begin{subfloat}[\label{CompExp_Xf20_w2_spect_exp}]
        {\includegraphics[width=0.4\textwidth]{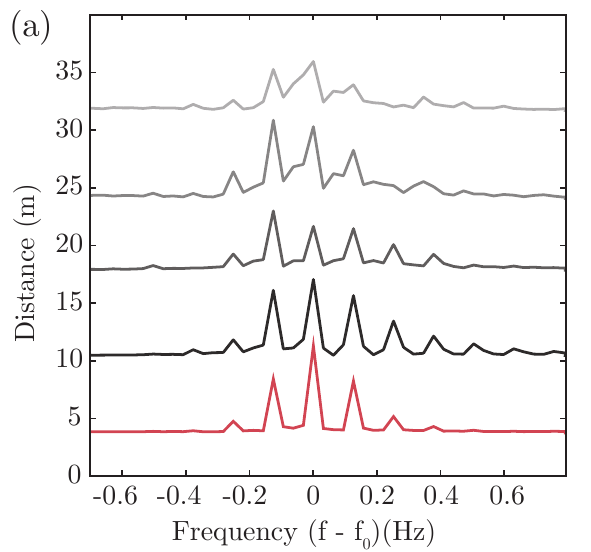}}     
    \end{subfloat}
    \begin{subfloat}[\label{CompExp_Xf20_w2_spect_sim}]
        {\includegraphics[width=0.4\textwidth]{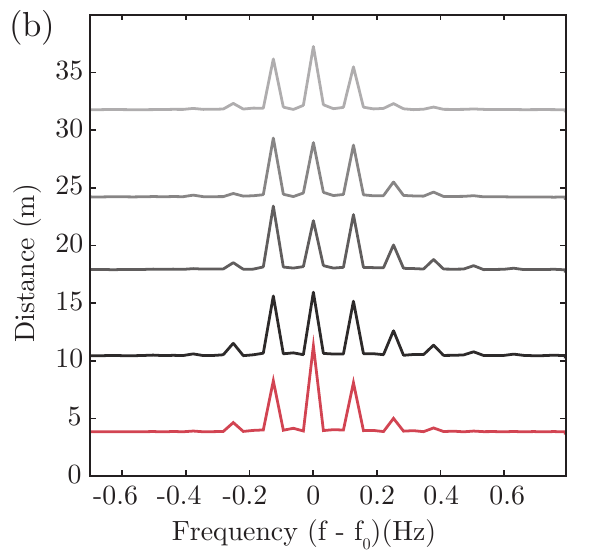}}      
    \end{subfloat}
    \begin{subfloat}[\label{CompExp_Xf20_w2_peakmean_exp}]
        {\includegraphics[width=0.4\textwidth]{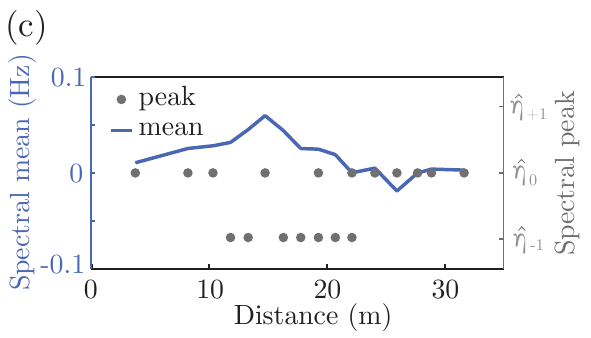}}    
    \end{subfloat}
    \begin{subfloat}[\label{CompExp_Xf20_w2_peakmean_sim}]
        {\includegraphics[width=0.4\textwidth]{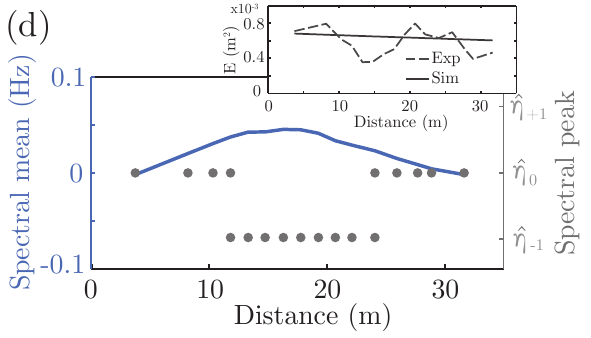}}     
    \end{subfloat}
     \caption{(a) Spectral amplitude $|\hat{\eta}|$ (normalized) offset by distance at $U$ = 2 m/s for (a) experiment and (b) simulations. Evolution of  $f_\textrm{m}$ (blue solid line) and $f_\textrm{p}$ (gray dots) for experiment (c) and simulation (d). For the same parameters as in figure \ref{fig_CompExp_Xf30_Spect}, except shorter focal distance $x^*=-20$~m. The inset shows the Energy evolution for the simulations and experiment}\label{fig_CompExp_Xf20_W2p0}
\end{figure}

Performing another experimental investigation for $U$ = 2 m/s and for a shorter expected focal distance of about 20 m from the wave maker ($x^*=-20$~m), as shown in figure \ref{fig_CompExp_Xf20_W2p0}, allows to quantify the spectral dynamics after the focal point. The spectral evolution indeed shows the same downshift behavior in the spectral peak, and a broad growth of the higher frequencies ($f>f_0$). However, for both of these features a reversion towards the initial condition occurs at the end of the tank. Indeed, for the homogeneous NLS, the Fermi-Pasta-Ulam recurrence \citep{Fermi1955} predicts a cyclic pattern after which the initial condition is completely recovered and then repeats several times. However, the MNLS framework predicts only a near-recovery \citep{Tulin1999,Lo1985}, that is, a quasi-recurrence. This is observed in figure 9 in \citet{Tulin1999} or figure 4 in \citet{Chabchoub2013a} without wind. With respect to the spectral mean, this quasi-recurrence causes an increase of the value of $f_\textrm{m}$ near the focal point and a decrease to the original value when the cycle is finished. As a consequence, the spectral mean `oscillates' if multiple quasi-recurrence cycles occur. The same oscillatory behavior is observed for the spectral peak as it shifts down and recovers. Together with the long range simulations in section \ref{sec_LongSim}, this reveals that the spectral peak downshift due to wind in the previous case, when the focal point was expected to be 30 m from the wave generator, is temporary too. As the wind forcing merely amplifies the asymmetry induced by the nonlinear Dysthe terms, so long as the wave does not break, the downshift of the peak will follow the oscillatory pattern set by these nonlinear Dysthe terms and is reversible.

\begin{figure}
    \centering
    \begin{subfloat}[\label{fig_CompExp_Xf30_w6_spect_exp}]
        {\includegraphics[width=0.40\textwidth]{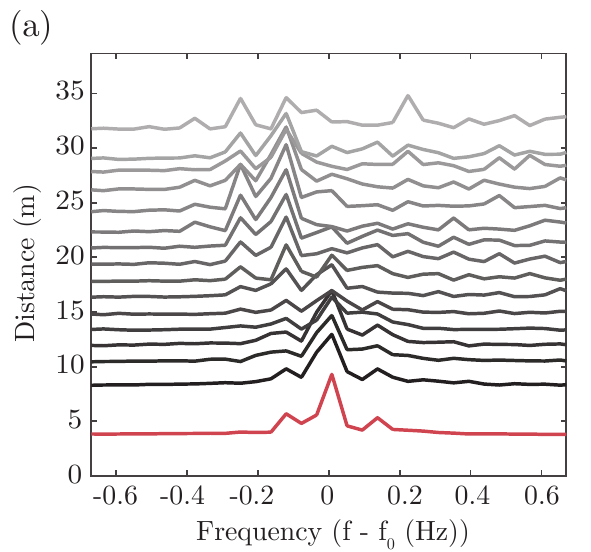}}  
    \end{subfloat}
    \begin{subfloat}[\label{fig_CompExp_Xf30_w6_spect_sim}]
        {\includegraphics[width=0.40\textwidth]{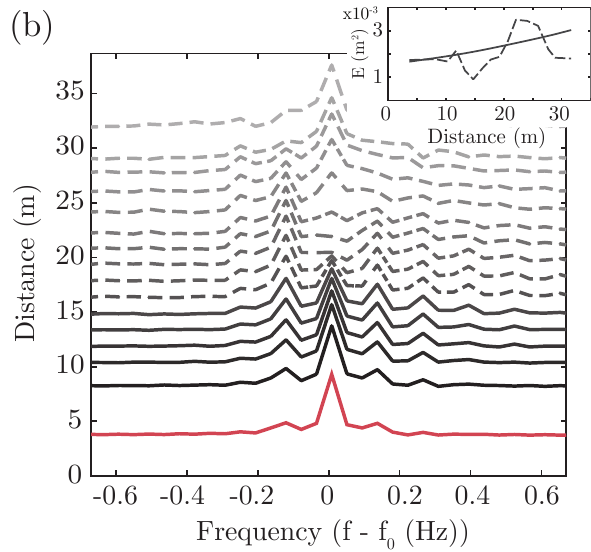}}
    \end{subfloat}
     \caption{Spectral amplitude $|\hat{\eta}|$ (normalized) offset by distance at $U$ = 6 m/s for (a) experiment and (b) simulation, for the same parameters as in figure \ref{fig_CompExp_Xf30_Spect}. The simulation is no longer accurate after the wave breaking event at $x \approx$ 15 m, indicated by the dashed lines. The inset shows the Energy evolution for the simulations and experiment. }\label{fig_CompExp_Xf30_w6_spect}
\end{figure}

\begin{figure}
    \centering
    \begin{subfloat}[\label{fig_Exp_wind_peak}]
        {\includegraphics[width=0.40\textwidth]{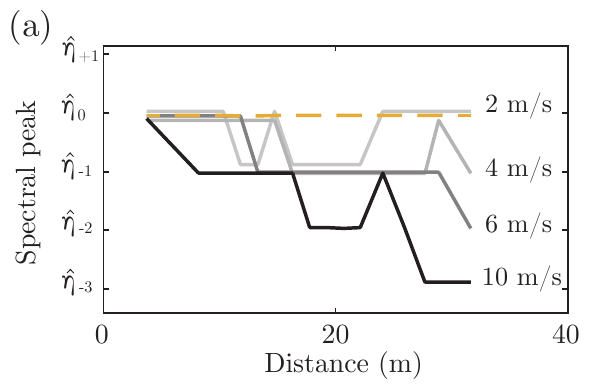}}
    \end{subfloat}
    \begin{subfloat}[\label{fig_Exp_wind_mean}]
        {\includegraphics[width=0.40\textwidth]{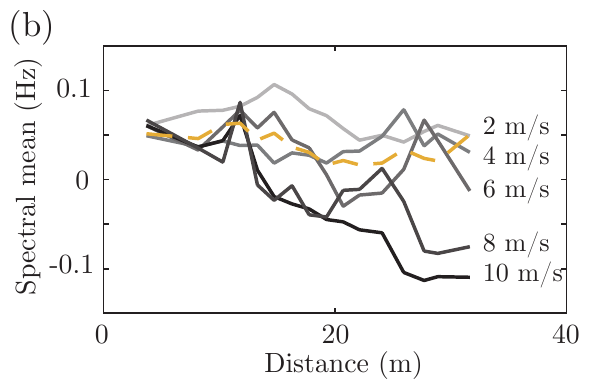}}
    \end{subfloat}
     \caption{Experimental evolution of (a) the spectral peak $f_\textrm{p}$ and (b) the spectral mean $f_\textrm{m}$  as a function of distance for increasing wind speeds, for same parameters as in figure \ref{fig_CompExp_Xf30_Spect}, except shorter focal distance $x^*=-20$~m.. A systematic downshift occurs for strong winds of (a) the spectral peak up to the position of $\hat{\eta}_{-3}$ and (b) the spectral mean. Dashed line indicates $U$ = 0 m/s. Wavebreaking  occurs for $ U~\geq~4$~m/s.}\label{fig_Exp_wind}
\end{figure}

The upper limit of the model, with respect to the steepness, is reached when wave breaking occurs. Figure \ref{fig_CompExp_Xf30_w6_spect} shows the spectral evolution for the expected focal point at 30~m from the wave maker ($x^*=-30$~m) and a stronger wind of $U$~=~6~m/s. Wave breaking has been observed to occur around $x~\approx~15$~m. Clearly, a downshift is observed in the experiments, from which $\hat{\eta}_0$ does not recover. In the simulations, this permanent downshift cannot be reproduced, instead, a quasi-recurrence occurred. This disagreement can be attributed to the wave breaking that is not considered in the model.

This observation underlines the importance of dissipation associated with wave breaking to induce a permanent downshift for both the spectral peak and mean. To further exemplify this fact, figure \ref{fig_Exp_wind_peak} shows the experimentally measured spectral peak shifts down several modes, up to $\hat{\eta}_{-3}$, for increasing wind speeds. A similar downward trend is observed for the spectral mean in figure \ref{fig_Exp_wind_mean}. For our carrier wave parameters, breaking occurs for $U>$~4~m/s, and therefore cannot be compared to simulations.

While the focus of this work lies on the forcing aspect of the wind, that is, up to the point of wave breaking, it is worth mentioning efforts to model the wave evolution after the point of wave breaking. \citet{Kato1995} propose an ad-hoc higher order term that activates at high steepness and leads to a permanent downshift of the peak. In a more theoretically structured approach, \citet{Trulsen1990,Trulsen1992} add a symmetric source term to the MNLS equation and observe a permanent downshift to the most unstable mode. However, applying this model to our data did not yield the permanent downshift observed in for instance figure \ref{fig_CompExp_Xf30_w6_spect_exp}. A rigorously derived model for the symmetric NLS is proposed and validated by experimental data in \citet{Tulin2002,Hwung2011}, where a downshift to the most unstable mode is also observed. Here, the asymmetry is caused by an integral over the envelope. This model seems a promising candidate to simulate wave breaking in the framework of an MNLS equation. 

As the NLS inherently only applies to narrow banded spectra, it remains an open question whether a wave breaking term in an NLS-like evolution equation can account for a downshift of multiple modes, and thus a broader spectrum, as observed in figure \ref{fig_Exp_wind_peak}. For fully nonlinear simulations, efforts to simulate wave breaking have been successfully conducted by \citet{Tian2010,Tian2011}.

In summary, we observe a temporary downshift of the spectral peak towards the lower satellite, and a temporary upshift of the mean. Our model (\ref{eqn_FullModel}) reproduces the qualitative features  of the experimental results well. These include the correct length scale and magnitude of the spectral mean and peak shift, the crossing of the sidebands, and the broadening of the spectrum. Considering the variability inherent to wind experiments, a qualitative agreement on the spectral dynamics is the best one can expect. To overcome the length limit of our wave tank and investigate multiple semi-recurrence cycles, long range simulations have been performed as described in the next section
 
% ======= Long Range Simulations ====== %
\section{Long range simulations} \label{sec_LongSim}
 
 Numerical simulations have been performed over a length of 100 m on AB with parameters $x^*=-20$~m, $A$~=~0.25 and carrier wave parameters $T_0$ = 0.6 s, $\eps$ = 0.1, $\nu = \SI{1e-5 }{m^2/s}$, $\Gamma = \SI{7.5e-3}{s^{-1}}$ under wind forcing, in which case $\Gamma > 5 \nu k_0^2$. To demonstrate the effect of our higher order wind term, three simulation cases are compared: 
\vspace{5mm}
\begin{enumerate}
  \item \hspace{5mm} \textit{Absence of wind}:  Equation (\ref{eqn_FullModel}) including the MNLS correction and both leading and higher order viscosity terms, however without wind. In this case, $\delta_0 < 0$ and $\delta_1 < 0$.
  \item \hspace{5mm} \textit{Leading order wind}: Equation (\ref{eqn_FullModel}) including the MNLS correction and both order viscosity terms, with only the leading order wind term, as posed in \citet{Trulsen1992,Kharif2010,Onorato2012}. In this case, $\delta_0 > 0$ and $\delta_1 < 0$. 
  \item \hspace{5mm} \textit{Full model}: Equation (\ref{eqn_FullModel}) including the MNLS correction and both order viscosity terms, with both the leading order and the higher order wind term. In this case, $\delta_0 > 0$ and $\delta_1 > 0$.
\end{enumerate}
\vspace{5mm}

% ---- FIGURE : LONG TANK Env + Spect ---- %
\begin{figure}[h!]
    \centering
    \begin{subfloat}[\label{fig_LongTank_ak10_env_NW}]
        {\includegraphics[width=0.40\textwidth]{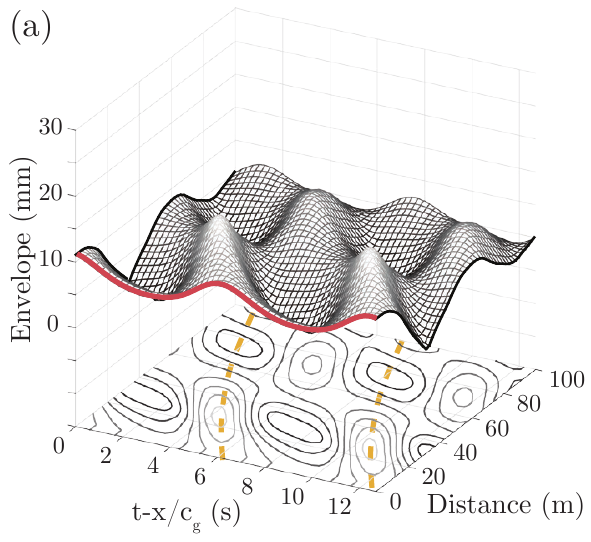}}       
    \end{subfloat}
    \begin{subfloat}[\label{fig_LongTank_ak10_env_WW}]
        {\includegraphics[width=0.40\textwidth]{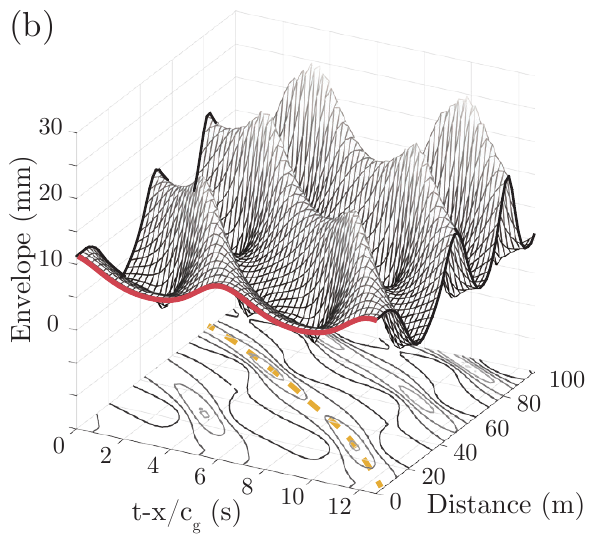}}        
    \end{subfloat}
    \begin{subfloat}[\label{fig_LongTank_ak10_env_SW}]
        {\includegraphics[width=0.40\textwidth]{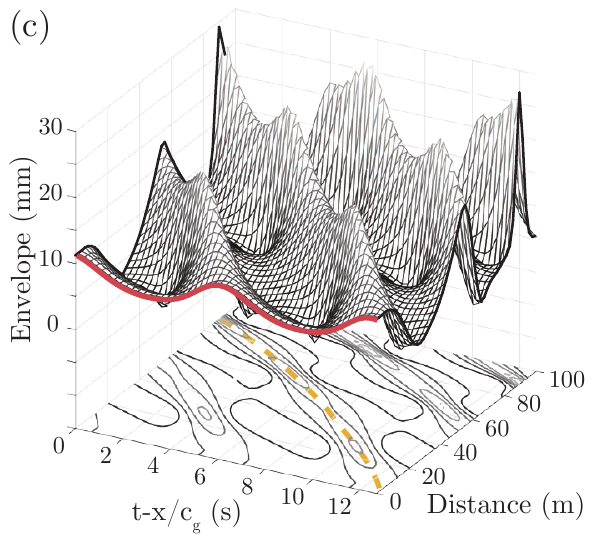}} 
    \end{subfloat}
    \begin{subfloat}[\label{fig_LongTank_ak10_spect_SW}]
        {\includegraphics[width=0.40\textwidth]{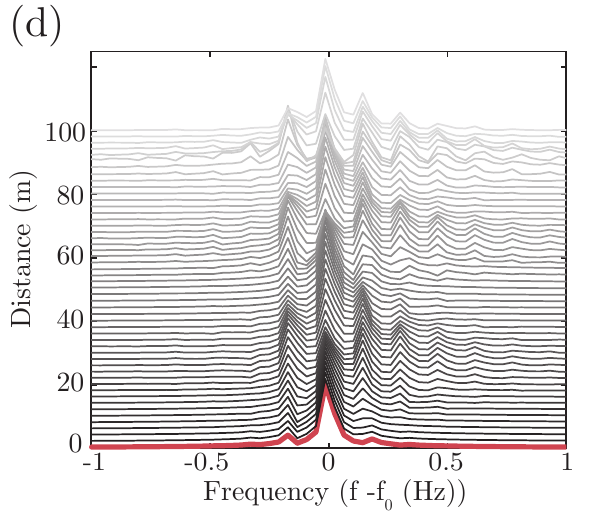}}       
    \end{subfloat}
     \caption{ Evolution of envelope in space and time for the (a) \textit{no wind}, (b)  \textit{leading order wind}, (c) \textit{full model} simulations. The orange dashed line shows the space-time evolution of the wave packets. (d) Simulated spectral amplitude $|\hat{\eta}|$ (normalized), for the \textit{full model} simulation. Each spectrum is offset according to the distance Red line indicates the initial condition based on the theoretical Akhmediev solution at $x^*=-20$~m and $A$ = 0.25 that is subsequently propagated in space by numerical integration. For carrier wave parameters $\eps$ = 0.1, $T_0= 0.6$~s, and simulation parameters $\nu~=~\SI{1e-5 }{m^2/s}$ and $\Gamma = \SI{7.5e-3}{s^{-1}}$ when wind is active.}\label{fig_LongTank_Env}
\end{figure}

Figure \ref{fig_LongTank_Env} compares the envelope amplitudes for these three cases. 
In absence of wind, simulation (i), the viscosity attenuates the amplitude. In addition, as described by \citet{Kimmoun2016}, it induces a shift in the quasi-recurrence pattern of the envelope, as indicated in figure \ref{fig_LongTank_ak10_env_NW}. Here, the dashed line is perpendicular on the gradient lines of the envelope amplitude. In simulation (ii), the wind amplifies the amplitude of the envelope with increasing fetch, as shown in figure \ref{fig_LongTank_ak10_env_WW}. In addition, the forcing cancels the shift caused by the viscosity term as now only maxima occur on the dashed line in figure \ref{fig_LongTank_ak10_env_WW}. In line with results reported in \citet{Kharif2010}, due to wind forcing and the increasing steepness, the position of the focal point is moved upstream to $x$~=~25~m, and every subsequent quasi-recurrence cycle is more compressed in space compared to the previous one. The result of simulation (iii) is similar to that of simulation (ii). The consideration of the additional higher order wind term causes a slight further increase of the envelope's amplitude, and shortens the length of the quasi-recurrence cycle: figure \ref{fig_LongTank_ak10_env_SW} now fits 3 maxima of the envelope. Figure \ref{fig_LongTank_ak10_spect_SW} displays the corresponding evolution of the normalized amplitude spectrum. Several MNLS quasi-recurrence cycles can be observed in which the spectrum broadens asymmetrically and narrows again. This quasi-recurrence pattern is superimposed on the general broadening of the spectrum due to wind action.

% ---- FIGURE : LONG TANK Spectral Parameters ---- %
\begin{figure}[h!]
	\centering
	\begin{subfloat}[\label{LT_meanpeak}]
     	{\includegraphics[width=0.38\textwidth]{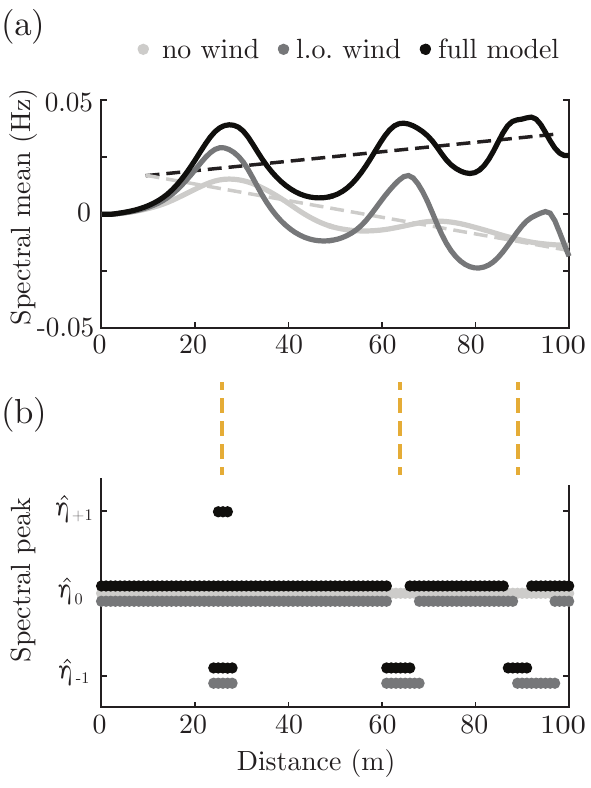}}     	
    \end{subfloat}
    \begin{subfloat}[\label{fig_LT_modes}]
      {\includegraphics[width=0.38\textwidth]{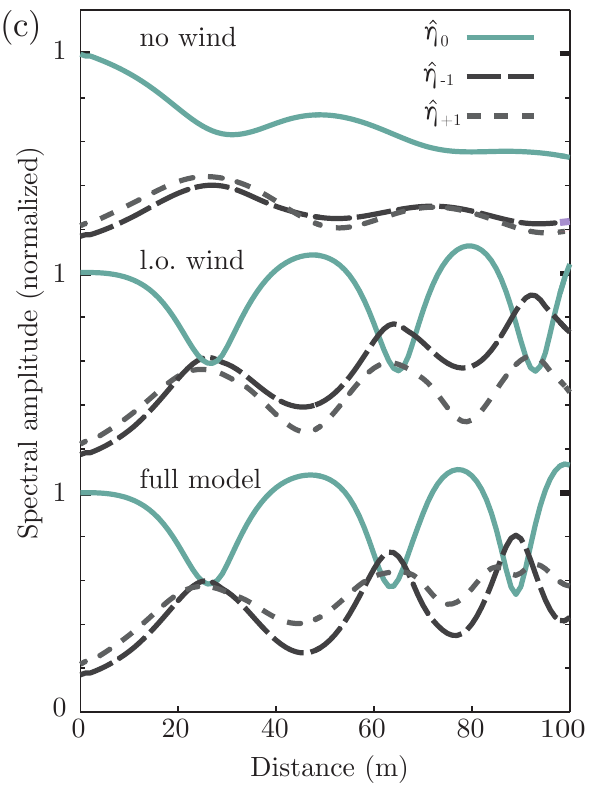}}     
    \end{subfloat}
  \caption{ (a) Simulated evolution of the spectral mean $f_\textrm{m}$. Dark dashed line is the trend set by the higher order wind term, light dashed line the trend set by the higher order viscosity term. (b) Evolution of the spectral peak $f_\textrm{p}$. Dots are offset for clarity. The dashed lines between (a) and (b) indicate the focal points of the quasi-recurrence cycles. (c) Simulated evolution of the Fourier amplitudes for of the three central modes $\hat{\eta}_0$ (green solid line), $\hat{\eta}_{-1}$ (dashed dark gray line) and $\hat{\eta}_{+1}$ (dashed light gray line). Comparing the \textit{no wind}, \textit{leading order (l.o.) wind} and \textit{full model} simulations.  Simulations based on the same parameters as figure \ref{fig_LongTank_Env}. }
  \label{LT_meanpeakmodes}
\end{figure}

Comparing the spectral evolution of the simulation cases, however, the influence of the higher order wind term can be clearly observed.
Figure \ref{LT_meanpeakmodes}a displays the spectral mean of the three simulations. Simulation (i) confirms the result of \citet{Carter2016}, namely that the higher order viscosity term causes a downshift in the spectral mean. Since $\Gamma =0$, $\delta_1 < 0$, and the light dashed line, indicating the trend of the spectral mean, has a negative slope. In simulation (ii), the addition of the leading order wind term accelerates the oscillation of the spectral mean, without affecting the trend: the oscillations follow the same downward slope as set by the higher order viscosity term ($\delta_1 < 0$). In contrast, in the full model simulation (iii), there is indeed a clear tendency towards a permanent upshift in the spectral mean, indicated by the dark dashed line, as now there is a forcing effect in the higher order ($\Gamma > 5 \nu k_0^2$, $\delta_1 > 0$). In brief, due to the MNLS modification terms the mean oscillates around the slope induced by a balance between the higher order viscosity term and the higher order wind term. 

Figure \ref{LT_meanpeakmodes}b shows the position of spectral peak $f_\textrm{p}$ as a function of distance. As observed in the experimental data in Section \ref{sec_ExpSimComp}, without wind forcing $f_\textrm{p}$ remains equal to $f_0$. In simulation (ii) a temporary downshift is observed. The origin of this downshift can be revealed by analyzing figure \ref{LT_meanpeakmodes}c. Without wind action, the amplitudes of the modes oscillate, but the sidebands do not overtake the main mode. Considering the leading order wind effect, the wave steepness is amplified and consequently the growth and decay rates of all modes, increasing the frequency and amplitude of their oscillation. In figure \ref{LT_meanpeakmodes}c the sidebands do overtake the main peak within a quasi-recurrence cycle. Due to the initially slightly different growth rates, $\hat{\eta}_{-1}$  reaches a slightly higher amplitude than $\hat{\eta}_{+1}$ and a temporary downshift in the spectral peak sense occurs. The observations of a similar spectral peak downshift pattern by \citet{Tulin1999} without wind forcing but at higher steepness, confirm that the spectral peak downshift is not a direct effect of wind forcing. Rather, it is an consequence due to the wind's influence on the steepness. That is, the spectral asymmetry inherent to the MNLS, where the lower sideband has a slightly higher growth rate than the upper sideband, is amplified by the wind as wave steepness is naturally increased. In simulation (iii) this downshift pattern is not significantly altered, although the quasi-recurrence cycles are slightly shorter. We can notice in figure \ref{LT_meanpeakmodes}c that instead of $\hat{\eta}_{-1}$ continuously being the dominant sideband, $\hat{\eta}_{-1}$ and $\hat{\eta}_{+1}$ alternate.

% ---- FIGURE : LONG TANK Group velocity + Steepness ---- %

\begin{figure}[h!]
\centering
\begin{subfloat}[\label{fig_LT_cg}]
        {\includegraphics[width=0.40\textwidth]{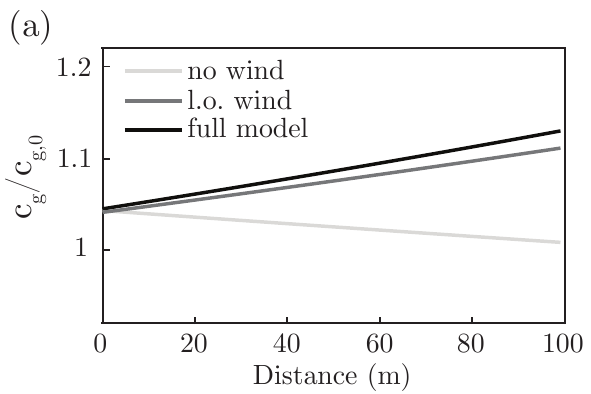}}
      \end{subfloat}
    \begin{subfloat}[\label{fig_LT_steepness}]
      {\includegraphics[width=0.40\textwidth]{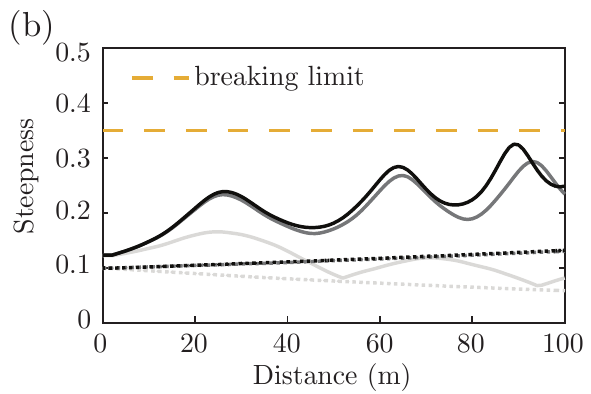}}     
    \end{subfloat}
  \label{fig_LT_cg_steepness}
  \caption{Simulated evolution of (a) the group velocity and (b) the steepness. The solid lines indicate the maximal steepness. The dotted lines indicate the characteristic steepness $\eps_\text{char}~=~\textrm{RMS}(\eta^2) k_0 $. Comparing the \textit{no wind}, \textit{leading order (l.o.) wind} and \textit{full model} simulations. The dashed line indicates the critical steepness for wave breaking. Simulations based on the same parameters as figure \ref{fig_LongTank_Env}. }
\end{figure}

The addition of the forcing terms affects the group velocity. The homogeneous NLS implies a linear group velocity $c_{g,0}$. Taking into account the MNLS correction terms increases the wave packets' speed \citep{Goullet2011}. The higher the steepness, the higher the importance of the MNLS correction and the higher the increase in $c_{g}$. For the \textit{no wind} simulation the dashed line in figure \ref{fig_LongTank_ak10_env_NW} shows a curvature of the direction of the wave packet propagation in the $x-t$ plane that is attenuated towards the end of the tank due to the decrease in steepness caused by viscosity and dissipation. In contrast, for the wind simulations, the dashed lines indicate that this induced variation of group velocity increases with the increase of the steepness. Figure \ref{fig_LT_cg} quantifies the deviation from the linear group velocity for the three simulation cases (i), (ii) and (iii). In the experiment, a similar increase in the group velocity is indeed observed.

It should be noted that the regime in which the higher order wind term becomes relevant is hard to reach experimentally due to its proximity to the wave breaking threshold. An increased steepness makes the higher order wind term of greater influence and hereby increases the deviation from simulation (ii). However, at the same time, high steepness brings waves closer to the breaking threshold, beyond which our model (\ref{eqn_FullModel}) is incomplete. Furthermore, the significance of the higher order wind term increases with the wind strength and fetch, see figure \ref{LT_meanpeakmodes}a. Indeed, the steepness is also increased as a consequence of wind strength and fetch, as displayed in figure \ref{fig_LT_steepness}. Thus it is important to monitor the steepness in the simulations to signal possible wave breaking. In our long range simulations, the maximal steepness  of the wave $\eps_\text{max} = k_0 a_\text{max}$, remains below the breaking limit of $\eps$ = 0.35 as considered by \citet{Trulsen1992}. While this value for the threshold number was calculated for Stokes waves in the absence of wind, \citet{Saket2017} show the breaking threshold is very similar for wind driven waves. Note that other studies suggest an even higher critical steepness \citep{Melville1982,Babanin2007, Toffoli2010}.

% ---- Table Results summary ---- %
\begin{table*}
\caption{\label{tab_summary}Our observations of the influence on the spectrum of the MNLS correction, and the leading order (l.o.) and higher order (h.o.) wind and viscosity terms in (\ref{eqn_FullModel}). US~=~upshift, DS~=downshift.}
\begin{ruledtabular}
\begin{tabular}{ccc}
 &\multicolumn{2}{c}{Effect}\\
 Terms & Mean & Peak  \\ \hline
MNLS correction   & temporary DS & temporary US \\
l.o. viscosity  $\rightarrow$ lower $\eps$  &  slower MNLS dynamics & slower,  damped oscillation   \\
h.o. viscosity  & - & permanent DS \\
l.o. wind  $\rightarrow$ higher $\eps$  & faster MNLS dynamics  & faster,amplified oscillation \\
h.o. wind  & - & permanent US \\
\end{tabular}
\end{ruledtabular}
\end{table*}

% ======= Discussion ====== %
\section{Discussion}  \label{sec_Discussion}

For the spectral peak, data and simulations alike show that a forcing wind can induce a downshift. However, the underlying cause for the faster growth of the lower sideband is the asymmetry introduced by the MNLS correction terms, which is amplified by the wind. Moreover, this downshift is only temporary. Considering the spectral mean, our long range simulations show that the higher order wind term creates a permanent upward trend, while the higher order viscosity term causes a permanent downward trend. As both terms have the same form in (\ref{eqn_FullModel}), the balance between these two, the sign of $\delta_1$, determines whether an upshift or a downshift in the spectral mean will be observed. Finally, when the wind action is sufficiently strong, wave breaking is a natural result. We experimentally confirm the well known notion that wave breaking induces a permanent downshift in both the spectral peak and spectral mean. Our observations on the effect of the wind, viscosity, and the MNLS modification are summarized in table \ref{tab_summary}. 

These findings might seem contradictory with respect to existing literature in which wind is often associated with spectral downshift of gravity waves \citep{Hara1991,Touboul2010,Schober2015,Kato1995,Trulsen1992}, as discussed. However, by taking into account the different downshift interpretations and making the distinction between wind forcing, in which energy is added to the system and forms of energy dissipation that can be triggered by wind, our results extend the existing framework, as upon closer inspection instances of permanent downshift are associated with dissipation. This distinction between the direct and potential indirects effects of wind solves the downshift paradox. This idea is confirmed in the review on frequency downshift by \citet{Dias1999}. Note that as pointed out in this review, the results in this work and the works discussed above are based on uni-directional waves, and the situation for directionally spread waves can be different \citep{Gibson2006}, as for instance demonstrated by \citet{Trulsen1997}.

The most obvious dissipative phenomenon that can occur as a result of wind forcing is that waves reach a critical steepness and break. Wave breaking shifts the spectral peak to a lower frequency. This has indeed been already experimentally observed by \citet{Lake1977,Melville1982,Tulin1999} and is explained along the general line of energy being dissipated  from the higher modes into the lower modes. Efforts have been made to model wave breaking theoretically  \citep{Trulsen1990,Trulsen1992,Kato1995,Tulin2002,Hwung2011,Tian2011}, as discussed. Another instance in which wind can have a dissipative effect is when it blows in opposite direction of the wave propagation and as such damps the waves. This configuration has been modeled by \citet{Schober2015}, a study in which the dissipation term as defined by \citet{Kato1995} has been taken into account and a permanent downshift is modeled as well. Finally, even when the direct forcing effect of wind is included in a study, the dominant regime for spectral movement can still be dissipative when the viscosity is strong. For example, \citet{Touboul2010} observed a permanent downshift in the spectral peak due to the effect of wind. However, they are exactly on the balance of forcing and dissipation in the leading order, $\Gamma \approx 4 k_0^2 \nu$, and thus, in the higher order regime the dissipation is slightly dominating, $\delta_1 < 0$. A similar argument holds for \citet{Hara1991}. While our study applies to surface gravity waves, it is interesting to note that for capillary-gravity waves \citet{Hara1994} numerically found a frequency upshift of the spectral peak due to wind, and \citet{Skandrani1996} have shown numerically that the shift to lower frequencies is promoted by a damping mechanism. 

% ======= CONCLUSION====== %
\section{Conclusion} 

We derive a higher order $O(\eps^4)$ wind forcing contribution to the MNLS framework, resulting in a forced-damped MNLS equation (\ref{eqn_FullModel}). The direct effect of this term, when it exceeds the viscosity at the same order, is an upshift of the spectral mean. This trend is superimposed on the oscillation caused by the MNLS correction. For significant wind action, the higher order wind term cancels the downshifting effect of the higher order viscosity term and moves towards an upward trend of the spectral mean. The leading order wind term is symmetric but can amplify the asymmetric growth initiated by the MNLS correction terms, resulting in a temporary downshift of the spectral peak. Finally, we confirm that the permanent downshift of the spectral mean and of the spectral peak often observed in wind experiments is an indirect effect associated with dissipation, including wave breaking.

This is the first time a propagation equation for deep water waves including wind forcing is validated by laboratory experiments. The evolution of an Akhmediev breather in the presence of wind shows good agreement with the model in the limited fetch, dictated by the length of the facility. A natural continuation of this work is an experimental exploration of the full damping-forcing range that we have now modeled in the leading and higher order in (2.8), in particular the upshift in the spectral mean predicted by the higher order wind term. In addition, increasing the fetch in the laboratory environment would improve the validation analysis of the numerical results. However, it is a delicate balance between observing the effect of the higher order term, and driving the model out of its validity range due to wave breaking, since the elements that increase the effect of the higher order term, {\it i.e.} fetch, wind speed and steepness, at the same time drive waves toward breaking.

\begin{acknowledgments}
We acknowledge financial support
from the Swiss National Science Foundation (project 200021-155970). The experiment was partly funded by the Labex MEC (Contract ANR-10-LABX-0092) and the A*MIDEX project (ANR-11-IDEX-0001-02). The authors thank Andrea Armaroli for fruitful discussions and Michel Moret for support on the computing infrastructure.
\end{acknowledgments}

% \nocite{*}
\bibliography{Bibliography}% Produces the bibliography via BibTeX.

\end{document}